\documentclass[aps,prx,twocolumn,showpacs,superscriptaddress,amsmath,amssymb]{revtex4-2}
\usepackage{graphicx}
\usepackage{latexsym}
\usepackage{bm} 
\usepackage{color}
\usepackage{epsfig}
\usepackage{multirow}
\usepackage{xcolor}
\usepackage{colortbl}
\usepackage{hhline}
\usepackage{simplewick}
\usepackage{soul}

\usepackage{graphicx}
\usepackage[colorlinks=true,linkcolor=blue,citecolor=blue,urlcolor=blue]{hyperref}
\usepackage{bbold}
\usepackage{gensymb}

\AtBeginDocument{%
    \newwrite\bibnotes
    \def\bibnotesext{Notes.bib}
    \immediate\openout\bibnotes=\jobname\bibnotesext
    \immediate\write\bibnotes{@CONTROL{REVTEX42Control}}
    \immediate\write\bibnotes{@CONTROL{%
    apsrev42Control,author="08",editor="1",pages="0",title="0",year="1"}}
     \if@filesw
     \immediate\write\@auxout{\string\citation{apsrev42Control}}%
    \fi
}%

\def \mbf {\mathbf}

\newcommand{\ket}[1]{|#1\rangle}
\newcommand{\braket}[3]{\langle#1|#2|#3\rangle}

\newcommand{\epvl}[1]{\langle#1\rangle}

\def \Tr {\mathrm{Tr}}
\newcommand{\comment}[1]{}

\definecolor{mygreen}{rgb}{0, 0.7, 0}

\begin{document}

\title{Odd-parity altermagnetism through sublattice currents: From Haldane-Hubbard model to general bipartite lattices}

\author{Yu-Ping Lin}
\affiliation{Department of Physics, University of California, Berkeley, California 94720, USA}
\author{Marc Vila}
\affiliation{Catalan Institute of Nanoscience and Nanotechnology - ICN2 (CSIC and BIST), Campus UAB, Bellaterra, E-08193, Barcelona, Spain}

\date{\today}

\begin{abstract}
We propose the sublattice currents in a compensated collinear magnetic system as a feasible route to odd-parity altermagnetism (ALM), where nonrelativistic collinear spin splitting occurs in the bands as an odd function of momentum. In contrast to previously classified ALMs, the sublattice currents break the time-reversal symmetry in the nonmagnetic crystal structure and allow for such odd-parity spin splitting. A representative example is the Haldane-Hubbard model at half filling. Although the compensated collinear magnetic ground state was previously recognized as antiferromagnetism, we show that it is actually an odd-parity ALM. Interestingly, its topological version serves as an example of an ALM Chern insulator. We further generalize the Haldane-Hubbard model to common two- and three-dimensional bipartite lattices. With spin splitting allowed by sublattice currents, the compensated collinear magnetic ground states at half filling are generally odd-parity ALM.
\end{abstract}

\maketitle

\addtocontents{toc}{\string\tocdepth@munge}

\textit{Introduction---} Recent studies of unconventional magnetism discovered altermagnetism (ALM) as the third type of collinear magnetic orders \cite{hayami19jpsj,smejkal20sa,yuan20prb,ma21nc,mazin21pnas,smejkal22prx1,smejkal22prx2,jungwirth24ax,dale24ax}. This compensated magnetic order resembles antiferromagnetism (AFM) for its zero net magnetization. Meanwhile, the lack of inversion-time-reversal \(\mathcal P\mathcal T\) symmetry breaks the Kramers degeneracy, leading to nonrelativistic spin splitting in the bands as ferromagnetism (FM). Under crystal symmetry breaking, the spin splitting manifests momentum-dependent alternating sign structures. With this unconventional combination of AFM and FM features, ALM is recognized as a promising platform for exciting phenomena, including innovative spintronics applications \cite{bai24afm}, anomalous Hall effects \cite{smejkal22nrm}, optical control of magnetization \cite{vila24ax,devita24ax}, and intricate interplay with superconductivity \cite{mazin25ab,papaj23prb,zhu23prb,brekke23prb,giil24prb,chakraborty24prb,bose24prb}. 

\begin{figure}[t]
\includegraphics[width=0.99\linewidth]{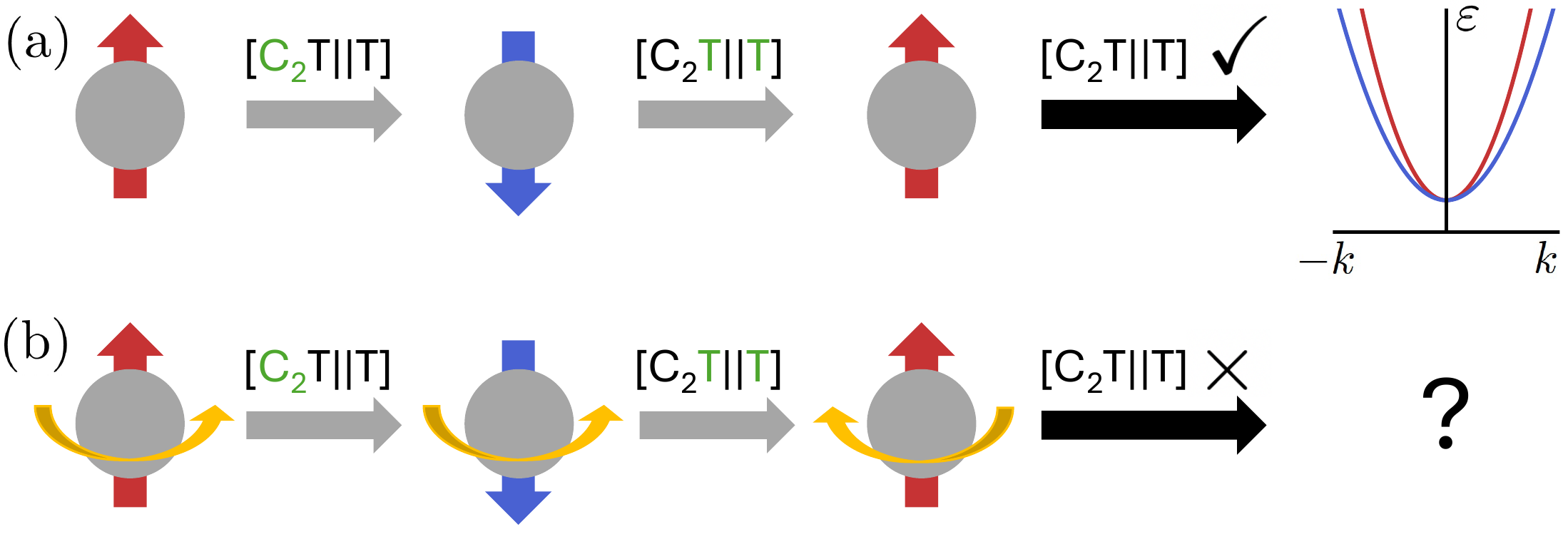}
\caption{\label{fig:F1} Schematic representation of the [$C_2\mathcal{T}||\mathcal{T}$] symmetry in a spinful system without (a) and with (b) time-reversal symmetry breaking in real space, as in the case of, e.g., orbital currents (orange arrow). When [$C_2\mathcal{T}||\mathcal{T}$] is preserved, the spin splitting is even in momentum. Red and blue arrows denote up and down spins, respectively. The green font shows what symmetry operation is applied.}
\end{figure}

In a previous classification with nonrelativistic spin groups \cite{smejkal22prx1}, it was assumed that collinear magnetic orders always have the spin group symmetry [$C_2\mathcal{T}||\mathcal{T}$], where the left (right) side of the bracket indicates a symmetry operation in the spin (orbital or lattice) space, with $C_2$ being a 180$\degree$ rotation perpendicular to the spin axis. This symmetry leaves the spin invariant since both $C_2$ and $\mathcal{T}$ flip the spin, as illustrated in Fig. \ref{fig:F1}(a). On the other hand, because the right-side $\mathcal{T}$ acts not only in real space but also in momentum space, it flips momentum ($k \rightarrow -k$). Therefore, the presence of such symmetry implies that the band structure follows $\varepsilon(k,s) \rightarrow \varepsilon(-k,s)$, that is, the spin splitting, if any, is even in momentum. However, there are situations where the [$C_2\mathcal{T}||\mathcal{T}$] symmetry is not preserved and consequently such even-parity splitting need not occur. While the breaking of the [$C_2\mathcal{T}|$ part occurs in noncoplanar systems \cite{birkhellenes23ax}, here we realize that the orbital or lattice symmetry $|\mathcal{T}$] can already be broken in systems with collinear magnetic order. For instance, Fig. \ref{fig:F1}(b) shows that, when orbital currents or magnetization are present, they break $\mathcal{T}$ in real space (i.e., spin independent), and hence [$C_2\mathcal{T}||\mathcal{T}$] is also broken. In that case, it is currently unknown how the spin splitting changes with momentum.

While nonrelativistic odd-parity spin splittings were explored in the phenomenology of spin Pomeranchuk instability \cite{wu04prl,wu07prb,kiselev17prb,wu18prb}, lattice models with spin-bond orders \cite{hirsch90prb,wu04prl}, or noncollinear magnetic orders \cite{birkhellenes23ax,brekke24prl,yu25prl}, the current classification of collinear magnetic systems does not contemplate such a scenario. Increasing the types of materials possessing ALMs with a different type of splitting would be important not only to broaden our classification and understanding of magnetic materials but also to enrich the potential spintornic applications brought about by spin splittings, such as magnetoresistance \cite{Shao2021, Smejkal2022TMR, Noh2025} or spin torques \cite{Gonzalez2021, Karube2022}. Altogether, a natural question arises: {\it Can we find odd-parity ALM, that is, odd-parity spin splitting in a compensated collinear magnetic system, by inducing time-reversal symmetry breaking in the nonmagnetic crystal structures?}

In this {\it Letter}, we answer this question by demonstrating repulsion-driven odd-parity ALM through sublattice currents. Unlike relativistic spin-orbit coupling \cite{sato24prl}, sublattice currents are nonrelativistic effects that break the time-reversal symmetry. A representative example of this scenario is the classic Haldane-Hubbard model at half filling \cite{he11prb,zheng15prb,wu16prb,arun16prb,vanhala16prl,imriska16prb,he24prb}. Previous studies discovered a compensated collinear magnetic ground state and classified it as an AFM. Although spin splitting in the bands was noticed \cite{zheng15prb,arun16prb}, its momentum-dependent symmetry structure was not studied. Here we revisit this spin splitting from the perspective of sublattice imbalance and spin group symmetry, thereby reclassifying the magnetic ground state as an odd-parity ALM. Interestingly, its topological version serves as an example of ALM Chern insulator. We further generalize the model to common two- and three-dimensional (2D and 3D, respectively) bipartite lattices, where odd-parity ALMs are consistently confirmed. Our work opens the avenue to the systematic search for odd-parity ALMs in correlated quantum systems, reveals sublattice currents as a mechanism for ALMs, and broadens our current classification of ALMs.

\begin{figure*}[t]
\centering
\includegraphics[scale=1]{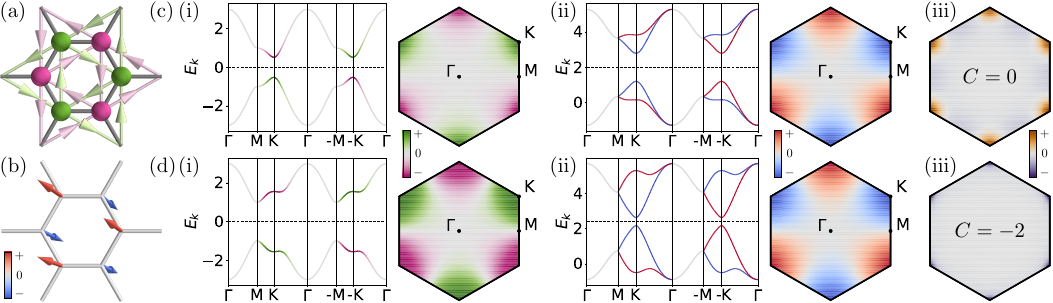}
\caption{\label{fig:haldane} Haldane-Hubbard model and repulsion-driven odd-parity ALM. (a) Haldane model on the honeycomb lattice (gray), where opposite currents flow on the second-neighbor bonds (lighter green and pink) in the two sublattices (green and pink). (b) Repulsion-driven ground state at half filling, where opposite FMs develop on the two sublattices. Here we show the on-site spin orders with colors representing \(z\) components. (c)  Haldane-Hubbard model with \(t_2=0.1\). (i) Left: sublattice imbalance \(w_{n\mbf k}\) occurs in the noninteracting bands. Right: BZ map of total sublattice imbalance \(w_{\mbf k}\) in the occupied bands. (ii) Spin splitting in the ALM at \(U_0=4\). Left: the bands split with nonzero spin polarization \(s_{n\mbf k}\). Right: BZ map of the spin-splitting energy in the occupied bands \(E^s_{\mbf k}\). (iii) BZ map of Berry curvatures in the ALM. (d) Haldane-Hubbard model with \(t_2=0.3\) and \(U_0=4.8\). The colors in (b)--(d) represent the respective data and follow the respective color bars.}
\end{figure*}

\textit{Haldane-Hubbard model---} We consider single-orbital models on 2D and 3D bipartite lattices, which are the minimal models with multisublattice structures. Our starting point is the classic Haldane-Hubbard model \cite{haldane1988}
\begin{equation}
\label{eq:haldane}
\begin{aligned}
H&=-t_1\sum_{\epvl{ij}_1\sigma}c_{i\sigma}^\dagger c_{j\sigma}-t_2\sum_{\epvl{ij}_2\sigma}c_{i\sigma}^\dagger e^{i\phi_{ij}}c_{j\sigma}\\
&\quad+\frac{1}{2}U_0\sum_{i\sigma\sigma'}c_{i\sigma}^\dagger c_{i\sigma'}^\dagger c_{i\sigma'} c_{i\sigma}
\end{aligned}
\end{equation}
on the 2D honeycomb lattice [Fig.~\ref{fig:haldane}(a)] at half filling. Here $c_{i\sigma}^{(\dagger)}$ annihilates (creates) a fermion at site \(i\) with spin \(\sigma=\uparrow\downarrow\); \(t_1=1\) and \(t_2=0.1\) represent the hoppings between first- and second-neighbor sites \(\epvl{ij}_{1,2}\), respectively; and \(U_0=4\) is the on-site repulsion. The second-neighbor hoppings carry complex phases \(\phi_{ij}=\pm\pi/2\), leading to uniform charge currents within each sublattice. Because of the opposite flows on the two sublattices, we refer to them as sublattice currents. Note that the elementary triangles in each sublattice host loop currents with three-1-in-1-out \((3+-)\) corners. These loop currents may be related to auxiliary staggered fluxes through the triangles, with zero net flux on the whole lattice \cite{haldane1988}. Using Hartree-Fock theory \cite{lin24prb,lin2411ax}, we obtain the interacting ground state on a finite-size (\(18^2\times2\)) lattice with a periodic boundary condition. With the on-site repulsion sufficiently large, the ground state manifests opposite FMs on the two sublattices [Fig.~\ref{fig:haldane}(b)]. This magnetic ground state remains robust under weak extended repulsion \cite{shao21prb,wang24prb}. In the previous literature \cite{he11prb,zheng15prb,wu16prb,arun16prb,vanhala16prl,imriska16prb,he24prb}, the magnetic order was recognized as an AFM. However, sublattice currents break the \(\mathcal{PT}\) symmetry and allow for spin splitting in the bands \cite{zheng15prb,arun16prb}. This important property suggests that the ground state is not an AFM in modern classification \cite{smejkal22prx1}.

As the central point of our work, we explain how the sublattice currents actually turn the ground state into an odd-parity ALM. We begin with the effect of sublattice currents in the noninteracting Haldane model. In the absence of sublattice currents with \(t_2=0\), the bands have equal weights in the two sublattices. This balance can be quantified by a zero sublattice weight
\begin{equation}
w_{n\mbf k}=\braket{u_{n\mbf k}}{\tau^3\sigma^0}{u_{n\mbf k}},
\end{equation}
where \(\ket{u_{n\mbf k}}\) is the \(n\)th-band Bloch state at momentum \(\mbf k\) and \(\tau^\mu=(\mathbb 1_2,\boldsymbol{\tau})\) with \(\boldsymbol{\tau}=(\tau^1,\tau^2,\tau^3)\) (same for \(\sigma^\mu\)) are the Pauli matrices in the sublattice (spin) sector. Since \(\omega_{n\mathbf{k}}\) measures a trivial identity \(\sigma^0\) in the spin space, it is basis independent in the bands with double spin degeneracy. At the Brillouin-zone (BZ) corners \(\pm\mbf K\), Dirac points with band crossings appear. The additional double degeneracy at these points can be understood from degenerate sublattice-polarized states \cite{lin2406ax}. When the sublattice currents are turned on with \(t_2=0.1\), the bands become sublattice-imbalanced \(w_{n\mbf k}\neq0\) [Fig.~\ref{fig:haldane}(c)(i)] \cite{castro23prl}. This imbalance is maximal at \(\pm\mbf K\), where the sublattice-polarized states are split. We further inspect the symmetry structure of this sublattice imbalance. The sublattice currents exhibit the momentum-space form factor
\begin{equation}
j_{\mbf k}=2t_2\sum_{m=1}^3\sin(\mbf k\cdot\mbf a_{2m})(\tau^3\sigma^0),
\end{equation}
where \(\mbf a_{2(m=1,2,3)}\) are the \(\text{C}_3\)-related Bravais-lattice unit vectors at second neighbors. The resulting sublattice imbalance is \(\text{C}_3\) symmetric, as confirmed by a BZ map of total sublattice weights in the occupied bands [Fig.~\ref{fig:haldane}(c)(i)]
\begin{equation}
w_{\mbf k}=\sum_{n=1}^2w_{n\mbf k}.
\end{equation}
Note that it takes an odd-parity form \(w_{\mbf k}=w_{-\mbf k}\) since inversion switches the sublattices.

We now turn on the on-site repulsion \(U_0=4\) and study the mean-field band structure \cite{lin24prb,lin2411ax} of the magnetic ground state. Under sublattice imbalance, the opposite sublattice FMs naturally drive spin splitting in the bands. Indeed, our computation finds band splitting with nonzero spin polarization [Fig.~\ref{fig:haldane}(c)(ii)]
\begin{equation}
s_{n\mbf k}=\braket{u_{n\mbf k}}{\tau^0\hat\sigma}{u_{n\mbf k}},
\end{equation}
where \(\hat\sigma=\boldsymbol{\hat s}\cdot\boldsymbol{\sigma}\) measures the spin polarization along the collinear magnetic direction \(\boldsymbol{\hat s}\). The spin splitting should inherit the odd-parity \(\text{C}_3\)-even structure from the sublattice imbalance. This symmetry structure is confirmed by a BZ map of the spin-splitting energy in the occupied bands [Fig.~\ref{fig:haldane}(c)(ii)]
\begin{equation}
E^\text{s}_{\mbf k}=\sum_{n=1}^2E_{n\mbf k}s_{n\mbf k},
\end{equation}
where \(E_{n\mbf k}\) is the mean-field dispersion energy. The discovery of this nonrelativistic spin splitting recognizes the ground state as an odd-parity ALM. Notably, its odd-parity nature goes beyond the previous classification \cite{smejkal22prx1}, which is enabled by the nonmagnetic time-reversal symmetry breaking from sublattice currents. The discovery of odd-parity ALM opens promising opportunities in spintronics: because the symmetry of the spin splitting is similar to that of spin-orbit-coupled split bands (i.e., odd in momentum), we expect similar phenomenology but with substantially larger magnitudes \cite{Chakraborty2025NC} as the splitting arises from exchange interaction rather than relativistic effects.

Although the noninteracting Haldane model is topological with the Chern number $C=-2$ \cite{haldane1988}, the ALM is topologically trivial with \(C=0\) under a sufficiently large repulsion [Fig.~\ref{fig:haldane}(iii)] \cite{he11prb,zheng15prb,wu16prb,arun16prb}. Interestingly, previous mean-field studies also found a topological regime with \(C=-2\) near the phase transition to normal insulator \cite{he11prb,zheng15prb,wu16prb,arun16prb}. The ground state in this regime serves as an example of ALM Chern insulator. We briefly explain these topological properties. In the noninteracting model, the negative Berry curvatures accumulate around \(\pm\mbf K\). In the strong ALM regime, where $U_0$ is large enough compared to $t_2$, a band inversion occurs between the spin-down (up) bands around sublattice-polarized \(\mbf K\) (\(-\mbf K\)) \cite{zheng15prb,arun16prb}. This process leads to positive Berry curvatures around \(\pm\mbf K\) in the occupied bands [Fig.~\ref{fig:haldane}(c)(iii)]. The remaining negative Berry curvatures migrate to the BZ edge centers \(\mbf M\)s and neutralize the newcoming positive ones. Therefore, the ground state becomes topologically trivial with \(C=0\). On the other hand, a weak ALM may not induce band inversion, thus keeping the ground state topological with \(C=-2\). We demonstrate this situation under stronger sublattice currents with \(t_2=0.3\), where the noninteracting band gap at \(\pm\mbf K\) becomes larger than the one at \(\mbf M\)s [Fig.~\ref{fig:haldane}(d)(i)] \cite{castro23prl}. Interestingly, the negative Berry curvatures now accumulate around \(\mbf M\)s, some of which migrate back to \(\pm\mbf K\) when ALM reduces the spin-down (up) band gap at \(\mbf K\) (\(-\mbf K\)). As ALM remains weak compared to the band gap, such as at \(U_0=4.8\) [Fig.~\ref{fig:haldane}(d)(ii)], the band inversion does not occur. The ground state remains a Chern insulator with negative Berry curvatures in the occupied bands [Fig.~\ref{fig:haldane}(d)(iii)].

\textit{General sublattice-current model---}Through the study of Haldane-Hubbard model, we have recognized the sublattice currents as a feasible route to odd-parity ALM. Building upon this knowledge, we now establish a systematic construction of lattice models that host repulsion-driven odd-parity ALM. Our construction generalizes the Hamiltonian (\ref{eq:haldane}) of the Haldane-Hubbard model with \(t_1 = 1\), \(t_2 = 0.1\), \(\phi_{ij} = \pm\pi / 2\), and \(U_0 = 4\) to the 2D and 3D bipartite lattices at half filling. This model introduces sublattice currents to the second-neighbor bonds, with uniform currents flowing oppositely in the two sublattices. Under sublattice currents, the noninteracting bands acquire sublattice imbalance. The strongest imbalance occurs at high-symmetry points, lines, or surfaces on the BZ boundary, where degenerate sublattice-polarized states \cite{lin2406ax} are split. Under on-site repulsion, the half-filling ground states naturally develop opposite sublattice FMs. With the aid of sublattice imbalance, this compensated collinear magnetic order induces nonrelativistic spin splitting in the bands and makes the ground states ALM.

\begin{figure*}[t]
\centering
\includegraphics[scale=1]{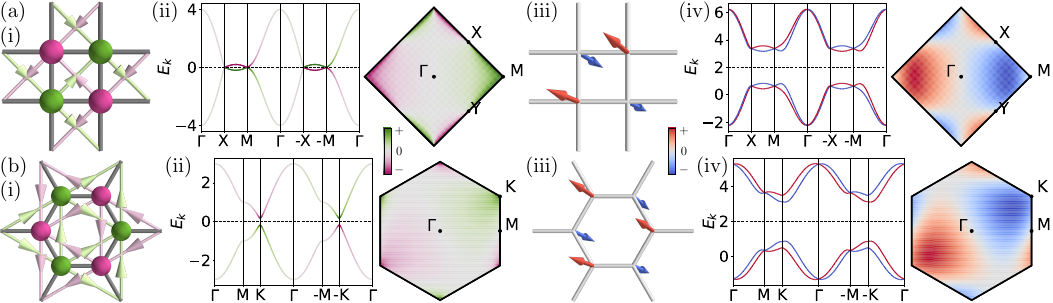}
\caption{\label{fig:2d} 2D bipartite-lattice models and repulsion-driven odd-parity ALMs. For (a) checkerboard and (b) honeycomb lattices, we show (i) model structures, (ii) sublattice imbalances in the noninteracting bands, (iii) opposite FMs in the two sublattices, and (iv) ALM-induced spin splittings in the bands. Note that (b) the honeycomb-lattice model exhibits a different sublattice-current configuration \((1++,1+-,1--)\) from the Haldane model. The formats of the figures follow Fig.~\ref{fig:haldane}.}
\end{figure*}

The sublattice currents exhibit the momentum-space form factor
\begin{equation}
\label{eq:slc}
j_{\mbf k}=2t_2\sum_{m=1}^{N_{2a}}\eta_m\sin(\mbf k\cdot\mbf a_{2m})(\tau^3\sigma^0).
\end{equation}
Here \(\mbf a_{2m}\)s and \(\eta_m=\pm1\) are \(N_{2a}\) Bravais-lattice unit vectors and current directions along different second-neighbor axes, respectively. Note that some lattices host multiple inequivalent configurations for sublattice currents. As we discuss later, different configurations support different types of spin splittings on the same lattice. Importantly, the ALMs in our systematic construction are all odd-parity, which expands the class of ALM beyond the previous classification \cite{smejkal22prx1}. This expansion is attributed to the nonmagnetic time-reversal symmetry breaking from the sublattice currents.

\textit{2D lattices---}We first investigate the bipartite lattices in 2D. As a starting point, we consider the simplest checkerboard lattice. There is only one inequivalent configuration for sublattice currents under lattice symmetry [Fig.~\ref{fig:2d}(a)(i)]. Note that the squares in each sublattice carry zero auxiliary fluxes. According to the form factor (\ref{eq:slc}) of sublattice currents with \(N_{2a}=2\), the sublattice imbalance manifests an odd-parity \(\text{R}_x\)-odd-\(\text{R}_y\)-even structure with maxima along the BZ boundary [Fig.~\ref{fig:2d}(a)(ii)]. Here \(\text{R}_{x,y}\) represent the reflections in the \(x\) and \(y\) directions, respectively. We confirm the repulsion-driven odd-parity ALM on a finite-size (\(16^2\times2\)) lattice [Fig.~\ref{fig:2d}(a)(iii)]. The band structure acquires spin splitting under ALM, following the same odd-parity structure as the sublattice imbalance [Fig.~\ref{fig:2d}(a)(iv)]. Notably, the ground state respects an effective time-reversal symmetry \(\mathcal{T} T_{x,y}\), where \(T_{x,y}\) are the one-site translations along the \(x\) and \(y\) directions, respectively. This symmetry enforces trivial topology in the ground state with zero Berry curvatures. We also note that such \(\mathcal{T} T_{x,y}\) symmetry is important to realize odd-parity magnetism in noncollinear systems \cite{birkhellenes23ax, brekke24prl}. However, it is not necessary in our collinear magnetic order where the lack of $[C_2 \mathcal{T} || \mathcal{T}]$ is what allows for the odd-parity spin splitting.

Although we have studied the honeycomb lattice in the Haldane-Hubbard model, there actually exists a second inequivalent configuration for the sublattice currents. By reversing the sublattice currents along one of the three second-neighbor directions, this configuration manifests the \((1++,1+-,1--)\) in-out corners in the elementary triangles of each sublattice [Fig.~\ref{fig:2d}(b)(i)]. The form factor (\ref{eq:slc}) with \(N_{2a}=3\) and \(\eta_1=-\eta_2=\eta_3=1\) breaks the \(\text{C}_3\) symmetry, leading to a twofold symmetric \(\text{R}_{\mbf K}\)-odd-\(\text{R}_{\perp\mbf K}\)-even sublattice imbalance [Fig.~\ref{fig:2d}(b)(ii)]. The repulsion-driven ALM [Fig.~\ref{fig:2d}(b)(iii)] then induces spin splitting in the bands with the same odd-parity structure [Fig.~\ref{fig:2d}(iv)]. Note that the noninteracting model also hosts a topological ground state with \(C=-2\), while the ALM trivializes the nontrivial topology with \(C=0\).

\begin{figure*}[t]
\centering
\includegraphics[scale=1]{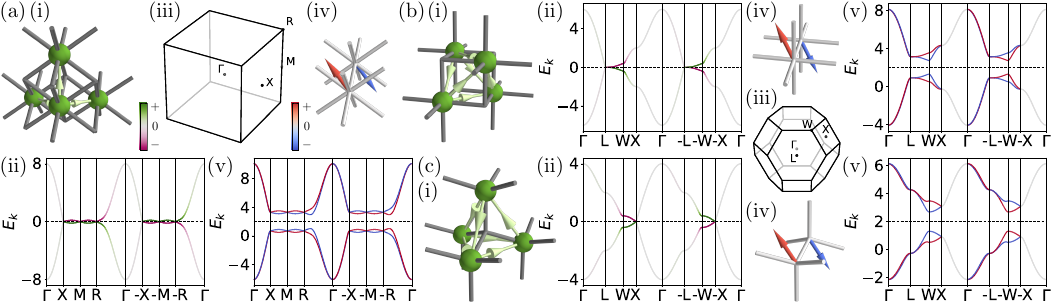}
\caption{\label{fig:3d} 3D bipartite-lattice models and repulsion-driven odd-parity ALMs. For (a) BCC, (b) 3D checkerboard, and (c) diamond lattices, we show (i) model structures, (ii) sublattice imbalances in the noninteracting bands, (iii) BZs, (iv) opposite FMs in the two sublattices, and (v) ALM-induced spin splittings in the bands. For (b) 3D checkerboard and (c) diamond lattices, we show the results for the sublattice-current configuration \((2++-,2+--)\). For clear illustration of (i) the model structures, we show the sites and currents only in one sublattice. The formats of the figures follow Fig.~\ref{fig:haldane}.}
\end{figure*}

\textit{3D lattices---}The Haldane-Hubbard model can also be generalized to 3D bipartite lattices. Here we consider the common examples of body-centered-cubic (BCC), 3D checkerboard, and diamond lattices (Fig.~\ref{fig:3d}). Our finite-size (\(8^3\times2\)) computation at half filling confirms the onset of repulsion-driven opposite FMs in the two sublattices. The spin splittings in the bands again recognize these compensated collinear magnetic ground states as odd-parity ALMs.

The BCC lattice stands on a simple-cubic Bravais lattice, which supports only one inequivalent configuration for sublattice currents [Fig.~\ref{fig:3d}(a)(i)]. Similar to the 2D checkerboard lattice, the sublattice imbalance is odd parity [Fig.~\ref{fig:3d}(a)(ii)] with maxima on the whole BZ boundary [Fig.~\ref{fig:3d}(a)(iii)]. The repulsion-driven ALM [Fig.~\ref{fig:3d}(a)(iv)] then induces spin splitting in the bands with the same odd-parity structure [Fig.~\ref{fig:3d}(a)(v)].

The 3D checkerboard and diamond lattices [Figs.~\ref{fig:3d}(b) and \ref{fig:3d}(c)] both stand on the face-centered-cubic (FCC) Bravais lattices. Because of their frustrated nature, the FCC Bravais lattices support more abundant structures for sublattice currents. There exist three inequivalent configurations with \((1+++,3+--)\), \((1+++,1++-,1+--,1---)\), and \((2++-,2+--)\) [Figs.~\ref{fig:3d}(b)(i) and \ref{fig:3d}(c)(i)] in-out corners in the elementary tetrahedrons of each sublattice. Different configurations exhibit the \(N_{2a}=6\) form factors (\ref{eq:slc}) with different symmetry structures. For example, the \((1+++,3+--)\) configuration obeys a \(\text{C}_3\) symmetry, which is absent in the other two configurations. The sublattice imbalances on these two lattices are again odd parity [Figs.~\ref{fig:3d}(b)(ii) and \ref{fig:3d}(c)(ii)], which are maximal along \(\mbf L\)-\(\mbf W\) and \(\mbf X\)-\(\mbf W\) lines on the BZ boundary [Figs.~\ref{fig:3d}(b)(iii) and \ref{fig:3d}(c)(iii)], respectively. The repulsion-driven ALMs [Figs.~\ref{fig:3d}(b)(iv) and \ref{fig:3d}(c)(iv)] inherit these symmetry structures and induce odd-parity spin splittings in the bands [Figs.~\ref{fig:3d}(b)(v) and \ref{fig:3d}(c)(v)].

Similar to the 2D checkerboard lattice, the ALMs on the BCC and 3D checkerboard lattices respect the effective time-reversal symmetry \(\mathcal{T} T_{1m}\). Here \(T_{1m}\) is the one-site translation along any first-neighbor directions. The discovery of these odd-parity ALMs in 3D can again be compared with recent studies of coplanar odd-parity magnetism \cite{birkhellenes23ax,brekke24prl,yu25prl}.

\textit{Discussion---} We have identified the sublattice currents in compensated collinear magnets as a feasible route to achieve odd-parity ALM by nonmagnetic time-reversal symmetry breaking. Starting from the Haldane-Hubbard model, we have established a systematic construction of 2D and 3D bipartite-lattice models, where on-site repulsion drives odd-parity ALMs at half filling. The search for practical materials with sublattice currents is presently an active line of research. While the kagome metals have been the most promising candidates for loop currents \cite{Liege2024, Gui2025, Fernandes2025, Chakraborty2025}, recent experiments have observed possible evidences for the Haldane-model-like sublattice currents in Mn\(_3\)Si\(_2\)Te\(_6\) \cite{zhang22n}. Meanwhile, ultracold atoms represent another attractive platform for the realization of odd-parity ALM \cite{das24prl}: they can realize Hubbard antiferromagnets \cite{mazurenko17n} as well as the Haldane model under Floquet drives \cite{jotzu14n}. While this work uses sublattice currents as the mechanism for odd-parity ALM, the same physics is expected in systems possessing orbital magnetization \cite{Liege2024, Gui2025, Chakraborty2025} or complex charge density waves \cite{Lin2021, Denner2021, Mielke2022}, as these effects also break time-reversal symmetry independently of the spin degree of freedom.

\textit{Note added---}We have become aware of follow-up work that investigated the spin-group analysis \cite{zeng26prb, luo25ax} and other possible origins of odd-parity ALM, such as Floquet drives \cite{Huang2026, Zhu2026, Li2026, Liu2026} and orbital orders \cite{zhuang25ax}.

\textit{Acknowledgments---}M.V thanks O.A. Ashour for helpful discussions. Y.-P.L. acknowledges the fellowship support from the Gordon and Betty Moore Foundation through the Emergent Phenomena in Quantum Systems (EPiQS) program. M.V. is supported by the Beatriu de Pin\'{o}s Grants Programme of the Department of Research and Universities, Government of Catalonia (2024 BP 00155). The ICN2 is funded by the CERCA Programme/Generalitat de Catalunya. The ICN2 is supported by the Severo Ochoa Centres of Excellence programme, Grant No. CEX2021-001214-S, funded by MCIN/AEI/10.13039.501100011033.




\bibliography{reference}

\section*{End Matter}
\label{endmatter}

\textit{Hartree-Fock theory---}In this section, we briefly introduce the Hartree-Fock theory for the study of interacting-fermion ground states. More pedagogical introductions can be found in Supplemental Material of Refs.~\cite{lin2411ax,lin24prb}.

Interacting fermions can be described by a general Hamiltonian
\begin{equation}
    H = \sum_{ab}\mathcal{H}_{0,ab}c_a^\dagger c_b + \frac{1}{2}\sum_{abcd}U_{acdb}c_a^\dagger c_c^\dagger c_d c_b.
\end{equation}
Here \(c_a^{(\dagger)}\) is the annihilation (creation) operator of a fermion in a basis state \(a\). The first and second terms represent the noninteracting and interacting parts of the model, respectively. We focus on the mean-field Slater-determinant states \(\ket{\Psi}\) of \(N\) fermions, where \(N\) effectively noninteracting states are occupied under Fermi statistics. This state can be represented by the density matrix
\begin{equation}
    P_{ab} = \epvl{c_b^\dagger c_a}
\end{equation}
with the projector condition \(P^2=P\). The energy of the state can be derived by the Wick's theorem and rewritten in a mean-field form
\begin{equation}
    E[P] = \frac{1}{2}\Tr[P(\mathcal{H}_0 + \mathcal{H}_\text{HF}[P])].
\end{equation}
Here the Hartree-Fock Hamiltonian
\begin{equation}
    \mathcal{H}_\text{HF}[P]_{ab}
    = \mathcal{H}_{0, ab} + \sum_{cd}(U_{acdb} - U_{acbd})P_{dc}
\end{equation}
captures an effectively noninteracting theory at the mean-field level
\begin{equation}
    H_\text{HF}[P] = \sum_{ab}\mathcal{H}_\text{HF}[P]_{ab}c_a^\dagger c_b.
\end{equation}

The mean-field ground state \(\ket{\Psi_\text{GS}}\) is obtained by minimizing the energy \(E[P]\) under the projector condition. This process yields the self-consistent equation for the ground-state density matrix \(P_\text{GS}\)
\begin{equation}
    [P_\text{GS}, \mathcal{H}_\text{HF}[P_\text{GS}]] = 0,
\end{equation}
under the condition that the ground-state density matrix \(P_\text{GS}\) is formed by the \(N\) lowest-energy eigenstates of the Hartree-Fock Hamiltonian \(\mathcal{H}_\text{HF}[P_\text{GS}]\). In practice, the numerical computation is achieved by an iterative algorithm. The algorithm begins with a random or designed density matrix. In each iteration, we obtain the Hartree-Fock Hamiltonian and update the density matrix from the lowest-energy eigenstates. The algorithm ends when the desired convergence of energy and density-matrix elements is achieved.

\textit{Magnetic order---}Once the ground-state density matrix is obtained, we can compute the on-site spin orders to understand the magnetic order of the ground state \cite{lin24prb}. In our lattice model, the basis index \(a = i\sigma\) contains the indices of site \(i\) and spin \(\sigma\). The computation of on-site spin orders is straightforward
\begin{equation}
    \mathbf{s}_i = \Tr\left(P_i\frac{\boldsymbol{\sigma}}{2}\right),
\end{equation}
where \(P_i\) is the \(2\times2\) on-site density matrix at site \(i\) and \(\boldsymbol{\sigma} = (\sigma^1, \sigma^2, \sigma^3)\) is the Pauli-matrix vector.

\textit{Band-structure analysis---}We can study the band structures of both the noninteracting Hamiltonian and the Hartree-Fock Hamiltonian \cite{lin2411ax}. In our study, the former tells us the effects of sublattice currents, while the later shows us the occurrence of odd-parity collinear spin splitting. Consider the noninteracting Hamiltonian
\begin{equation}
    \tilde{H} = \sum_{\tilde{i}\tilde{i}' \tilde{\tau}\tilde{\tau}'}\sum_{\sigma\sigma'}\tilde{\mathcal{H}}_{\tilde{i}\tilde{i}' \tilde{\tau}\tilde{\tau}'\sigma\sigma'}c_{\tilde{i}\tilde{\tau}\sigma}^\dagger c_{\tilde{i}'\tilde{\tau}'\sigma'},
\end{equation}
which represents either the noninteracting Hamiltonian or the Hartree-Fock Hamiltonian. Here \(\tilde{i}\) is the Bravais-lattice-site index under the periodicity of the Hamiltonian, \(\tilde{\tau}\) is the sublattice index within the unit cell \(\tilde{i}\), and \(\sigma\) is the spin index. We map the Hamiltonian to momentum space by the Fourier transform
\begin{equation}
    c_{\tilde{i}\tilde{\tau}\sigma} = \frac{1}{\tilde{N}_\text{BL}^{1/2}}\sum_{\mathbf{k}}c_{\mathbf{k}\tilde{\tau}\sigma}e^{i\mathbf{k}\cdot\mathbf{r}_{\tilde{i}\tilde{\tau}}},
\end{equation}
where \(\tilde{N}_\text{BL}\) is the number of Bravais lattice sites, \(\mathbf{k}\) is the momentum, and \(\mathbf{r}_{\tilde{i}\tilde{\tau}}\) is the position of the lattice site \(\tilde{i}\tilde{\tau}\). The mapping yields the momentum-space representation of the Hamiltonian
\begin{equation}
    \tilde{H} = \sum_{\mathbf{k}\tilde{\tau}\tilde{\tau}'}\sum_{\sigma\sigma'}\tilde{\mathcal{H}}_{\mathbf{k}\tilde{\tau}\tilde{\tau}'\sigma\sigma'}c_{\mathbf{k}\tilde{\tau}\sigma}^\dagger c_{\mathbf{k}\tilde{\tau}'\sigma'}
\end{equation}
with the Bloch Hamiltonian
\begin{equation}
    \tilde{\mathcal{H}}_{\mathbf{k}\tilde{\tau}\tilde{\tau}'\sigma\sigma'} = \sum_{\tilde{i}'}\tilde{\mathcal{H}}_{\tilde{0}\tilde{i}'\tilde{\tau}\tilde{\tau}'\sigma\sigma'}e^{-i\mathbf{k}\cdot(\mathbf{r}_{\tilde{0}\tilde{\tau}} - \mathbf{r}_{\tilde{i}'\tilde{\tau}'})}.
\end{equation}
Diagonalizing the Bloch Hamiltonian \(\tilde{\mathcal{H}}_{\mathbf{k}}\) at each momentum \(\mathbf{k}\), we get the \(n\)th-band Bloch states \(\ket{u_{n\mathbf{k}}}\) and their dispersion energies \(E_{n\mathbf{k}}\).

\end{document}